\documentclass[preprint]{aastex}  % preprint1 produces a single-column, single-spaced document
\usepackage[varg]{txfonts}
\usepackage{amssymb,wasysym}
\usepackage{graphicx, natbib}
\bibpunct{(}{)}{;}{a}{}{,}

\shorttitle{Photoionization rates for helium: update}
\shortauthors{Sok{\'o}{\l} and Bzowski}

\begin{document}

\title{\LARGE Photoionization rates for helium: update}

\author{Justyna M. Sok{\'o}{\l} \and Maciej Bzowski}
\affil{Space Research Centre of the Polish Academy of Sciences, Warsaw, Poland, \email{jsokol@cbk.waw.pl}}

%Abstract
\begin{abstract}
The NIS~He gas has been observed at a few AU to the Sun almost from the beginning of the space age. To model its flow an estimate of the loss rates due to ionization by solar extreme-ultraviolet (EUV) flux is needed. The EUV irradiance has been measured directly from mid 1990-ties, but with high temporal and spectral resolution only from 2002. Beforehand only EUV proxies are available. A new method of reconstruction of the Carrington rotation averaged photoionization rates for neutral interstellar helium (NIS~He) in the ecliptic at 1~AU to the Sun before 2002 is presented. We investigate the relation between the solar rotation averaged time series of the ionization rates for NIS~He at 1~AU derived from TIMED measurements of EUV irradiance and the solar 10.7~cm flux (F10.7) only.  We perform a weighted iterative fit of a nonlinear model to data split into sectors. The obtained formula allows to reconstruct the solar rotation averages of photoionization rates for He between $\sim$1947 and 2002 with an uncertainty ranging from less than $10\%$ during solar minimum up to $20\%$  for solar maximum.
\end{abstract}

\keywords{Solar Irradiance: Observations, Photoionization rates: model}

\section{Introduction}
Neutral interstellar helium (NIS~He) is an important indicator of the conditions in the local interstellar medium surrounding the Sun. It penetrates the heliosphere almost unaffected, its flow is modified inside the heliopause only by the gravitational attraction of the Sun and losses due to ionization. Solar radiation in extreme-ultraviolet (EUV) is the most important source of ionization of helium atoms in the heliosphere. Observations of the inflow of neutral interstellar gas have been carried out almost since the beginning of the space age. Recently \citet{frisch_etal:13a} published a compilation of all available observations and measurements of NIS~He from 1972 and found that the direction of the inflow may be changing with time. For analysis of measurements done for many years, like for a homogeneous analysis of Ulysses/GAS measurements of NIS~He from 1994 to 2007 \citep{bzowski_etal:14a} or pick-up ions analysis over the solar cycle \citep{chen_etal:13a} a continuous, homogeneous in time, and free of calibration issues series of ionization losses has to be known.

The solar EUV irradiance has been measured almost continuously from 1996, when SOHO (\emph{Solar and Heliospheric Observatory}) was launched with the \emph{Charge Element, and Isotope Analysis System / Solar EUV Monitor} (CELIAS/SEM) \citep{judge_etal:98}. Since then, two missions have started to monitor the solar EUV flux with high resolution in time and wavelengths: \emph{Thermosphere Ionosphere Mesosphere Energetics Dynamic} (TIMED, \citet{woods_etal:05a}) in 2002 and \emph{Solar Dynamics Observatory} (SDO, \citet{woods_etal:12a}) in 2010. Before the direct measurements of solar irradiance became available, the proxies of solar EUV flux were used to model the ionization rates. The most common EUV proxies are the 10.7~cm solar radio flux (F10.7, \citet{tapping:87}) and the solar magnesium II core-to-wing irradiance ratio (MgII index, \citet{heath_schlesinger:86}). Both are long-term data series, the first available from $\sim 1947$ and the second from $\sim 1978$. Other EUV proxies widely discussed in the literature (e.g. \citet{tobiska_etal:08a, dudokdewit_etal:09a}) are mainly composites of these two (often as a linear dependence); also models of solar spectrum are constructed with gaps filled by these two commonly used EUV proxies (e.g. \citet{chamberlin_etal:07a, tobiska_etal:00c}). The EUV proxies free of modelling issues should be used to derive photoionization rates before direct measurements of EUV flux to avoid additional uncertainty or bias of the constructed model.

\citet{bzowski_etal:13b} and \citet{bochsler_etal:14a} published models of reconstruction of the photoionization rates for neutral interstellar atoms (including He) in the heliosphere based on direct measurements of the solar EUV flux and a selection of proxies most relevant for photoionization rates studies. \citet{bochsler_etal:14a} used direct measurements of the solar EUV flux from \emph{Solar Extreme Ultraviolet Experiment} (SEE) onboard TIMED (Level~3A, version~11) and SOHO/CELIAS/SEM first order flux (26-34~nm), the MgII~core-to-wing ratio from SOLSTICE/SORCE, and the F10.7~cm flux provided by National Oceanic and Atmospheric Administration (NOAA). \citet{bzowski_etal:13b} based their model on a limited set of TIMED/SEE (Level~3A, version~11) data with the reference series from SOHO/CELIAS/SEM from both wavebands (first (26-34~nm) and central order (0.1-50~nm) flux), the composite MgII index from LASP (Laboratory for Atmospheric and Space Physics), and the F10.7~index from NOAA. Both models cover more than 3 full solar cycles (SCs) and end just after the minimum between solar cycles 23/24. 

Recently \citet{wieman_etal:14a} published a paper about an important correction needed in the \\ SOHO/CELIAS/SEM data and \citet{snow_etal:14a} informed on an improvement in the MgII~c/w ratio data from SOLSTICE/SORCE. In the light of these papers, the models proposed by \citet{bochsler_etal:14a} and \citet{bzowski_etal:13b} need to be updated, because the models based on the previous version of SEM and MgII SORCE data with shown inaccuracies should not be used. \citet{bochsler_etal:14a} and \citet{bzowski_etal:13b} discussed the EUV proxies relevant for modeling of photoionization rates, they considered three the best at the time of writing: SEM, MgII core-to-wing index, and F10.7. But, before the new, corrected sets of SEM and MgII data will be available, the selection of EUV proxies to reconstruct the photoionization rates before direct measurements has to be revised, because all models and EUV indices related to these two (like S10.7 or M10.7 from \citet{tobiska_etal:08a}) and developed before papers by \citet{wieman_etal:14a} and \citet{snow_etal:14a} may be affected by the issues announced in the two latter papers. In this paper we present a reconstruction of the photoionization rates for He from $\sim 1947$ based on the third proxy pointed by \citet{bochsler_etal:14a} and \citet{bzowski_etal:13b}, which seems to be free of systematic issues, it is F10.7.

F10.7 is the only one long-term, calibration-homogeneous, continuously measured, and publicly available series of EUV proxy available presently. As discussed by \citet{chen_etal:11a, chen_etal:12a}, F10.7 is not perfect proxy to retrieve the EUV flux on short time scales, there are F10.7-related proxies like S10.7 or M10.7 \citep{tobiska_etal:08a} that seem to better retrieve solar irradiance, but they are developed from the SEM and MgII series, respectively, before 2009 and may be affected by the recently published issues in these series. If the SEM response function and the reference solar spectrum used in the processing of SEM data need correction, all quantites based on this data have to be revised for reliability. When corrected data from SEM and MgII~c/w become available, using them to retrieve the ionizaion rates before the TIMED epoch will have to be reconsidered. Currently, in lack of reliable longterm series of SEM and MgII data, F10.7 seems to be the only available proxy relevant for the study of ionization rates before direct measurements of EUV irradiance. We do not use other available composite series of MgII$_{\mathrm{c/w}}$ ratio like those described in \citet{snow_etal:14a}, because they are composed of many short time series of measurements and the gaps are usually filled using correlation with F10.7 index, what means that this series is not independent of other proxies.

The reconstruction of photoionization rates for He presented in this paper is based on the solar rotation\footnote{1 siderial solar rotation = 27.2753 days (Carrington rotation)} averages of in-ecliptic photoionization rates derived from TIMED and F10.7 flux, both adjusted to 1~AU from the Sun. This time resolution and proxy selection is sufficient for NIS~He studies, as discussed earlier by \citet{bzowski_etal:13b, bochsler_etal:14a}, but regrettably is not enough for terrestrial studies.

\section{Data sources}
\subsection{TIMED/SEE}
TIMED began science operations at the beginning of 2002, launched on a circular orbit at 625-kilometers from Earth. The SEE instrument measures the solar spectral irradiance from 0.5 to 194.5~nm in 1~nm intervals \citep{woods_etal:05a}. For normal operations, SEE observes the Sun for about 3 minutes every orbit ($\sim 97$ minutes), which usually gives 14-15 measurements per day. Accuracy of daily measurments ranging from $10-20\%$ \citep{woods_etal:05a}. A suborbital (sounding rocket) payload is flown approximately once a year to help maintain TIMED/SEE absolute calibrations. Also three SDO/EVE calibration flights were used for checking the SEE degradation trends. The absolute calibration of the TIMED spectra irradiance is described in detail by \citet{woods_etal:05a}.

SEE data are available\footnote{Version~11 available at http://lasp.colorado.edu/see/see\_data.html} in two data products dedicated to study the ionization processes: Level~3 and Level~3A. The SEE Level~3 data are time-averaged over the entire day, while Level~3A is a product with a higher resolution in time and it contains irradiances from each orbit. Level~3 data are corrected for atmospheric absorption and instrument degradation. A contribution of flares is removed (what is recommended for long-term climatological study), and the flux is scaled to the distance of 1~AU from the Sun. They are, in our opinion, the most suitable for the goal of calculations of photoionization in the heliosphere at timescales greater than days. Level~3A contains the same corrections as Level~3, but does not remove flares, so is useful for short-term studies. We checked that the difference between Carrington rotation averaged series with and without flares affect the calculated photoionization rates for helium by less than $1\%$, what is negligible compared with other uncertainties in the proposed model.

The TIMED measurements are unique as they span the most puzzling time of the recent cycles of solar activity. TIMED was launched just after the maximum of solar cycle 23 and observed the lowest and longest minimum of solar activity  during the space age. Now it is operating in the maximum phase of SC~24, which is one of the weakest of the previously recorded maxima from the beginning of space age. The phase of decrease of SC~23 was prolonged, the increasing phase of SC~24 was shorter, and the current maximum is lower. Before the comparison of the minimum of SCs 22/23 and 23/24 based mainly on the SEM measurements \citep[e.g.][]{didkovsky_etal:10a} it has been assumed that the level of EUV flux at solar minimum does not vary from one solar cycle to another, just as the F10.7 flux. Now this question is still open. Partial verification of this assumption will be possible after the SEM measurements, which cover the last two solar minima, are corrected \citep{wieman_etal:14a}. Given these considerations, the TIMED measurements from 2002 to 2014 may not be directly comparable with any other previously measured series of EUV irradiance.

\subsection{F10.7 index}
The F10.7 index is an indicator of solar emission at the wavelength of 10.7~cm from sources present on the solar disk. It is a measure of the solar flux density in a band centered at 2.8~GHz. It is expressed in solar flux units (1~sfu = $10^{-22}$ W m$^{-2}$ Hz$^{-1}$). The flux in this waveband is a sum of three components (rapidly-varying, slowly-varying, quiet), coming up due to different processes, which may be differently distributed over solar surface and which may vary with time independently. The observations of the total emission in 10.7~cm are performed by ground base radio telescope three times per day during specific hours, so the data series are daily observations for a given hours in a day, not daily averages. The current status of the solar 10.7~cm flux from radio observations curated by DRAO (Dominion Radio Astrophysical Observatory, Canada) is described in details by \citet{tapping:13a}. This time series of F10.7 has been continuously available from 1947, what makes them a good proxy for modeling of the past photoionization rates of helium. The accuracy of F10.7 values is 1~sfu or $1\%$ of the flux value, whichever is the larger \citep{tapping:13a}. In our study we use the time series of the daily noon time flux adjusted to 1~AU, released by NOAA\footnote{Available at ftp://ftp.ngdc.noaa.gov/STP/space-weather/solar-data/solar-features/solar-radio/}.

\section{Model of the past photoionization rates based on F10.7}
\label{sec:phHeModel}
We reconstruct a longtime series of Carrington rotation averages of photoionization rates for helium based on the TIMED/SEE direct measurements of the EUV flux from 2002 and the F10.7 index for years before TIMED launch. Our aim is to construct long, continuous, and homogeneous series of Carrington rotation averaged photoionization rates for He to cover the whole time span of observations of neutral intrestellar helium in the heliosphere, i.e., from the beginning of 1970-ties until present, starting with use of simple methods, easy to implement by everyone using publicly available data series. The model is suitable to reconstruct Carrington rotation averages of the ionization rates, but its accuracy in reproducing variations on a finer time scale is limite, because the F10.7 and the solar EUV flux are not perfectly correlated during solar rotation period. As qualitatively shown by \citet{bochsler_etal:14a}, the phase shift between F10.7 and EUV irradiance may be significant and changing from one rotation to another. However, the correlation between solar rotation averages is generally very good for the photoionization rates for He, as we show in this paper.

We derived our model under the assumption that TIMED measurements are free from unaccounted degradation, long-term related issues, and from systematic wavelength-dependent imperfections. If there is an unaccounted systematic shift in the measured flux, the rates derived from TIMED and the analysis based on them will be affected in a systematic way. The reconstruction of photoionization rates from proxy is made under the assumption that the relation between the EUV flux and the proxy over the solar cycle does not change with time. The obvious requirement of invariable correlation between a proxy and the reconstructed quantity is common to all proxies.

In the first step of construction of the model of photoionization rates for He we calculated the ionization rates by integration of direct measurements of the daily solar flux from TIMED/SEE/Level~3 (version~11) data using the following formula:
\begin{equation}
\beta_{ph}\left(t\right) = \int\limits_{\lambda{_1}}^{\lambda{_2}}{F\left(\lambda,t \right)\sigma \left( \lambda \right) \mathrm{d}\lambda}
\label{eq:betaPhDef}
\end{equation}
where $(t,\lambda)$ is time and wavelength, respectively, $\sigma \left( \lambda \right)$ is the cross section for ionization by photons from \citet{verner_etal:96}, $\lambda_1 = 0.5$~nm (boundary of the spectrum measured by TIMED) and $\lambda_2 = 50.42$~nm (first ionization threshold for He), $F\left(\lambda,t \right)$ is the measured spectral irradiance. Next we average the daily photoionization rates over one solar rotation.

The currently available data are point measurements at the location of the detector. To correctly model the photoionization rates in the heliosphere on time scales shorter than one solar rotation, the full 3D information about the solar irradiance has to be available. In lack of this information available, we decided to calculate the average photoionization rates for He over one solar rotation.

Having the Carrington rotation averages of the photoionization rates for He, we reconstruct the rates before 2002 using solely the F10.7 proxy. We average the daily series of F10.7 over one solar rotation, similarly as we do with the photoionization series. In our case, the most important source of uncertainty of the photoionization is the spread of the daily rates covering the time range used to calculate the monthly average. We calculate the uncertainty of the Carrington rotation average from the standard deviation of the daily sets.

\begin{figure}[h!]
	\begin{center}
	\includegraphics[scale=0.6]{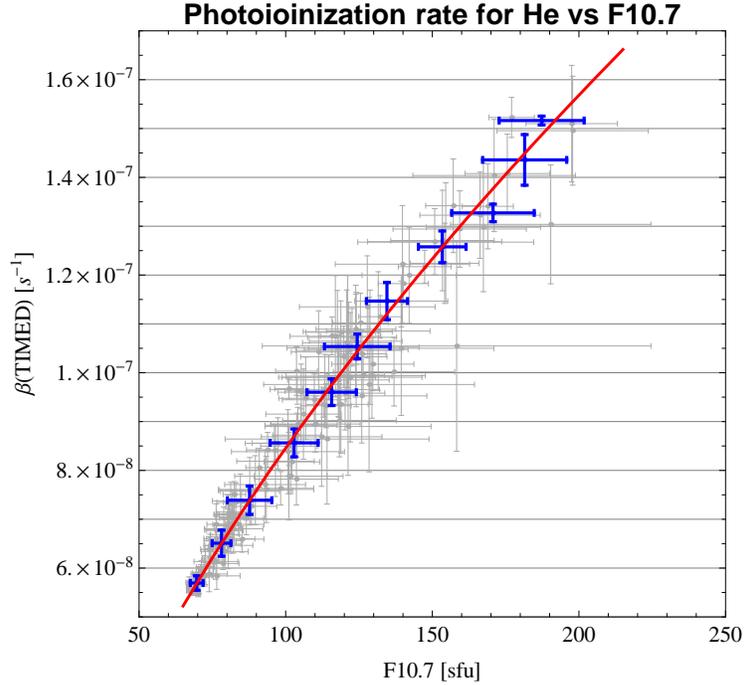}
	\end{center}
	\caption{Nonlinear fit from Equation~\ref{eq:final} (red line) to the Carrington rotation averages of photoionization rates for He at 1~AU from the Sun, obtained from direct integration of TIMED measurements of spectral irradiance and F10.7~cm flux observations (gray dots) for time span from the beginning of science operations of TIMED until Carrington rotation 2147.5. The same set sectored over photoionization values in shown by blue points. Error bars for both kinds of points (gray and blue) are standard deviations of the set account for average value, daily data and data inside sector, respectively. The horizontal grid lines mark the boundaries of the sectors.}
	\label{figF107vsPhTIMEDv2}
\end{figure}

The relation between the solar rotation-averaged values of photoionization rates for He and the F10.7 flux for the years of TIMED operation is quite good, as shown by gray points with errors bars in Figure~\ref{figF107vsPhTIMEDv2}. The considered time interval covers the decreasing phase of SC~23, the deep prolonged minimum, and the increasing phase of SC~24. There are more points with low values of F10.7 and photoionization rate than points with greater values, as manifested by the tight cluster of points in the lower left corner of Figure~\ref{figF107vsPhTIMEDv2}. This nonuniform distribution of points along the correlation line may bias the fitted function. To avoid this unwanted effect we divide the whole set into sectors and calculate an average value of photoionization rate and F10.7 inside each sector. The sectors are defined according to the photoionization rate values, with a constant step of $1 \times 10^{-8}$~s$^{-1}$, from $2 \times 10^{-8}$ to $2 \times 10^{-7}$. The sectored data set is presented by blue points with error bars in Figure~\ref{figF107vsPhTIMEDv2} and the boundaries of the sectors are shown by the horizontal grid lines. The errors bars are again calculated from the standard deviation of the set for each sector, as we are interested in the scatter of the points inside each sector.

It is clear from Figure~\ref{figF107vsPhTIMEDv2} that the points do not lay around a straight line. We fit a nonlinear function, $y=ax^p+b$, to the sectored data set to get the functional relation to calculate the predicted photoionization rate for He based on the F10.7 value. We do not want to miss the information about the scatter of the points taken to calculate the sectored data and we fit a relation with the uncertainties taken into account. We use an iterative procedure of fitting of nonlinear function to the data with errors in both coordinates, which corresponds to fitting with weights \citep[e.g.][]{fasano_vio:88a, press_etal:07a}. The fitting was performed using the \textit{Mathematica~8} package (procedure \texttt{NonlinearModelFit}) with weights defined in the following way:
\begin{equation}
w^{1}_i=\frac{1}{\sigma_{y_i}^2+\sigma_{x_i}^2}
\end{equation}
in the first step of iteration and:
\begin{equation}
w^{n}_i=\frac{1}{\sigma_{y_i}^2+(apx^{p-1}\sigma_{x_i})^2}
\end{equation}
in the next $n$ steps of iteration, where the number of iterations $n \geq 2$; $\sigma_{y_i}$ is the standard deviation of the $y_i$ values (in our case the photoionization rates) and $\sigma_{x_i}$ is the standard deviation of the $x_i$ values (F10.7 here). The iteration process stops when the average ratio of the parameters determined in step $n+1$ to step $n$ is smaller than $10^{-15}$. The resulting formula to calculate the solar rotation averaged time series of photoionization rates for He from F10.7 is the following:
\begin{equation}
\beta_{\mathrm{He}} = A~(F10.7)^{P} + B
\label{eq:final}
\end{equation}
with 
\begin{itemize}
\item[]$A=1.043037324829111\times10^{-8},$
\item[]$B=-6.28639375227517\times10^{-8},$
\item[]$P=0.5751941168680867.$
\end{itemize}
We give the coefficients of the fitted model with all digits to allow to reconstruct the model with the full precision, but the digits in the coefficients can be rounded to 4 digits after decimal point, not fewer, to keep the precision of reconstruction better than $1\permil$. We do not recommend further rounding of the number of digits in this formula because the reconstruction may change. We stress that the coefficients of the derived model do not have a physical interpretation, they are only a numerical representation of the relation between the investigated data sets. 

\begin{figure}
	\begin{tabular}{cc}
	\includegraphics[scale=0.53]{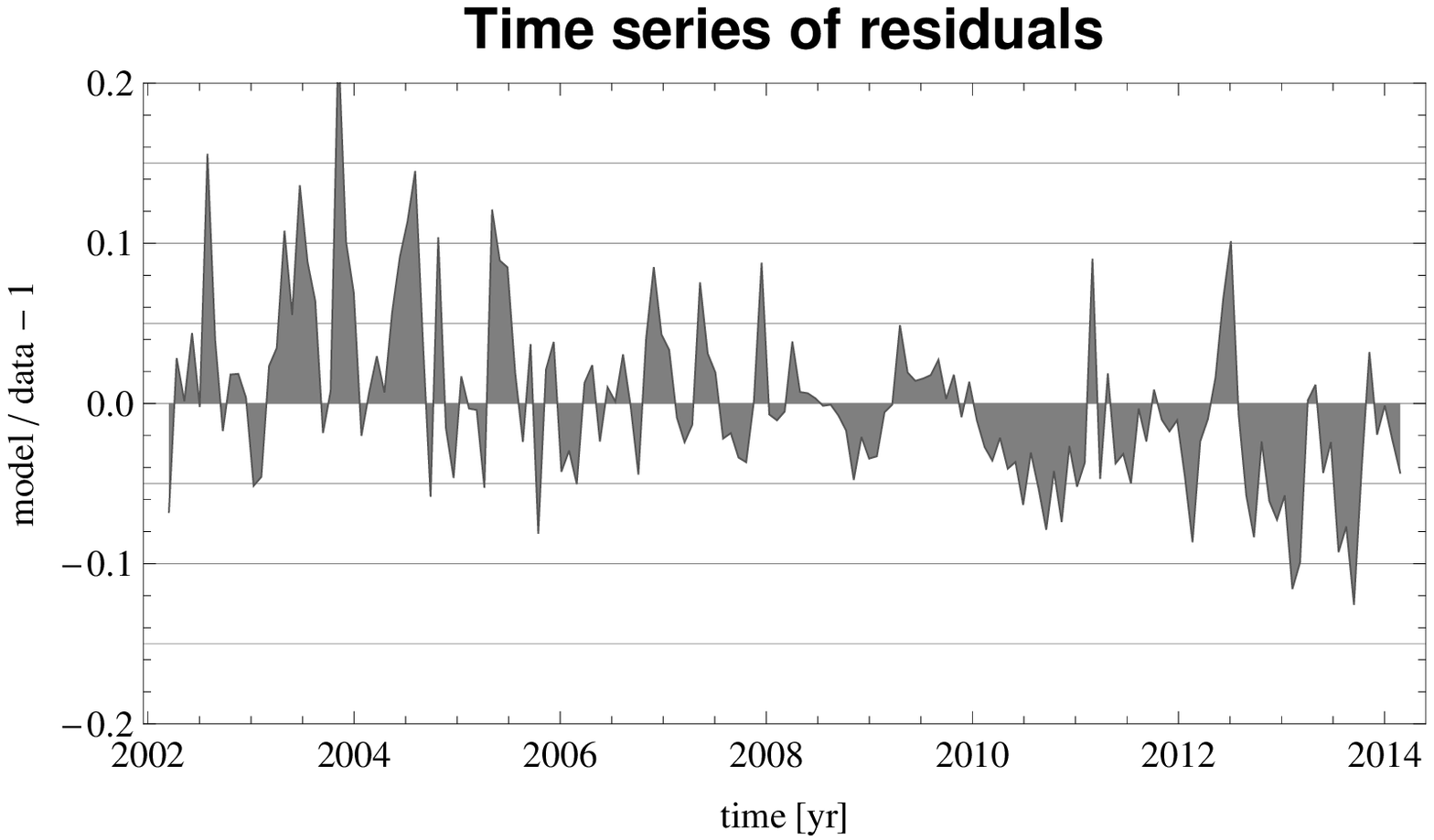} & \includegraphics[scale=0.6]{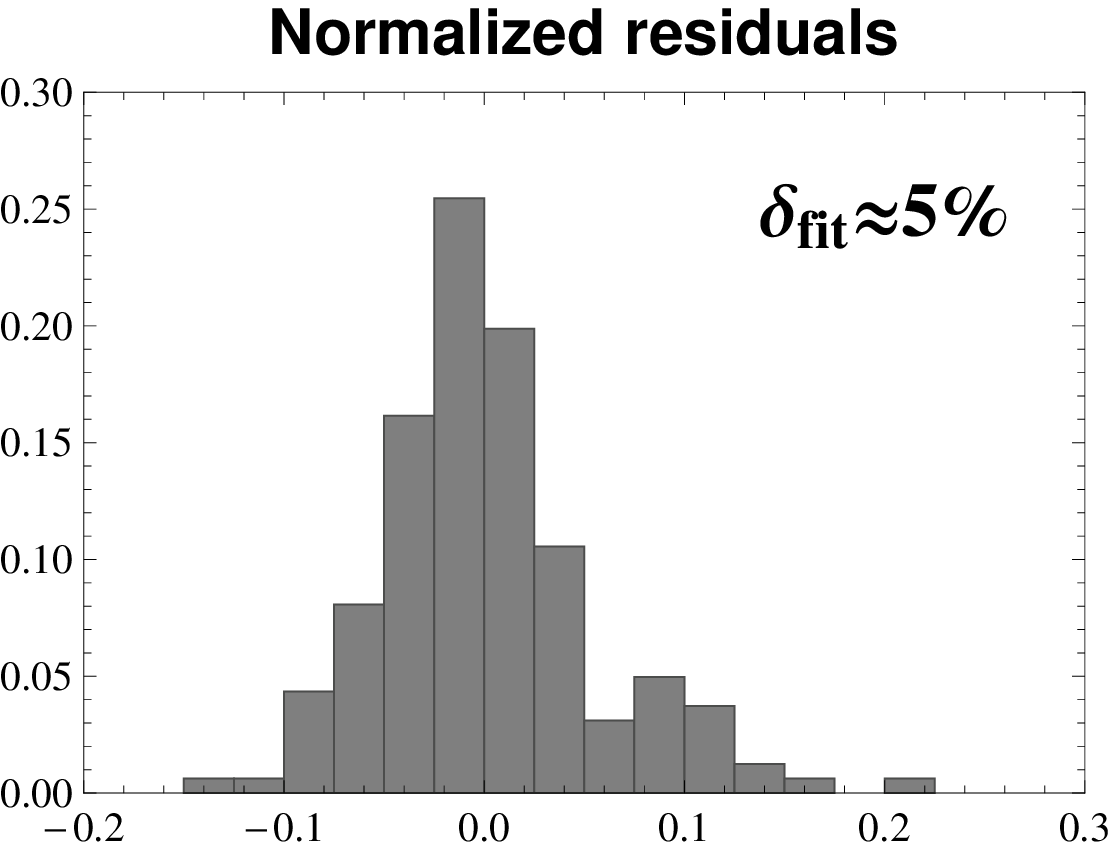} \\
	\end{tabular}
	\caption{Residuals (model / data - 1) of the model from Equation~\ref{eq:final} to the photoionization rates for He derived from TIMED at 1~AU for Carrington rotation averaged time series from 2002 to 2014. Upper panel: time series of the residuals. Lower panel: normalized histogram of residuals.}
	\label{figResiduals}
\end{figure}

The uncertainty of the model is assessed from the histogram of the residuals, shown in the lower panel of Figure~\ref{figResiduals}. The one-sigma accuracy of the model is $\sim5\%$. The residuals shown as time series in the upper panel of Figure~\ref{figResiduals} feature some spikes, which are due to spikes in the F10.7 Carrington rotation averaged data. The residuals are almost normally distributed, with a bump around $0.1$, which is mostly due to the spikes from the periods of high solar activity. The reconstruction is the best for the solar minimum conditions and the discrepancies between the model and the data increase with an increase of solar activity. The change in the residuals' trend before and after $\sim2010$ may indicate a time dependence of the relation between photoionization rates for He and F10.7, which is beyond the assumptions of the presented model. To verify this, we must wait until the next solar minimum, when the comparison of measurements from two minima will be possible. We note, however, that the correlation between most of the commonly used proxies seem to feature some cycle-to-cycle related variations, as can be seen in Figure~5 in \citet{tobiska_etal:08b}.

\begin{figure*}
	\resizebox{\hsize}{!}{\includegraphics{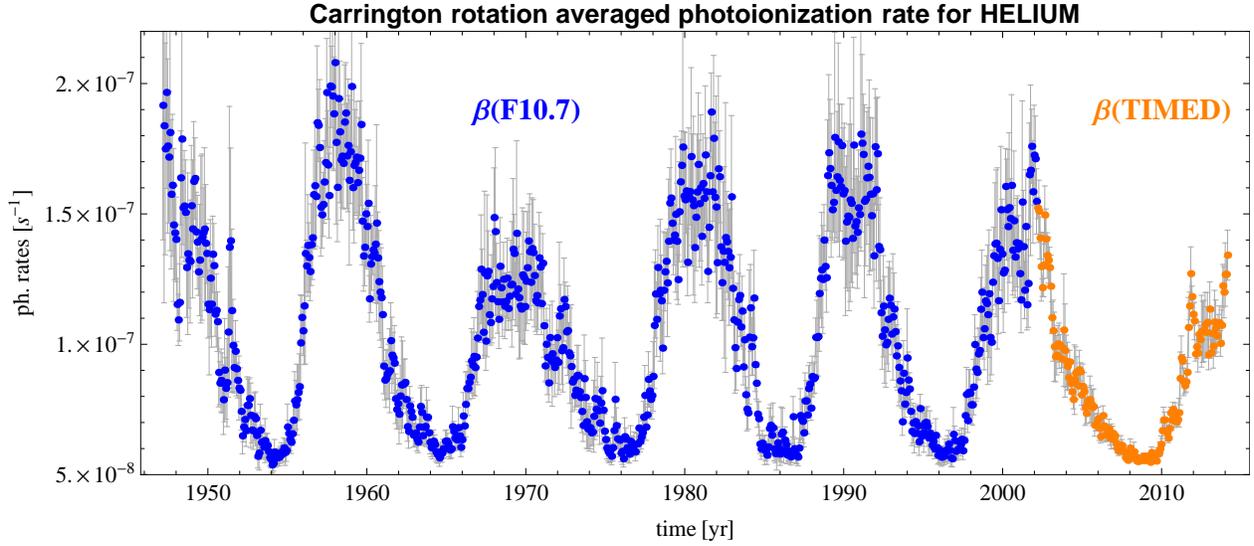}}
	\caption{Photoionization rates for helium at 1~AU in the ecliptic plane based on direct measurements of EUV flux from TIMED (orange points) and the F10.7 proxy (blue points) with Carrington rotation averaged resolution in time. The error bars (for clarity plotted in gray) are the uncertainties of the solar rotation averages of photoionization rates derived from TIMED and uncertainties from Equation~\ref{eq:DeltaPhModelSem} for the rates calculated from F10.7 according to formula given in Equation~\ref{eq:final}.}
	\label{figHeSeries5}
\end{figure*}

The final Carrington rotation averaged time series of the photoionization rates for helium (from direct integration of TIMED measurements and reconstructed from F10.7) is presented in Figure~\ref{figHeSeries5}.
The accuracy of the photoionization calculated from the integration of the directly measured EUV irradiance depends on a number of factors (see the detailed discussion in \citet{bzowski_etal:13b}). In general, the following effects account for it: the accuracy of the cross section for the photoionization (it is about $5\%$ for He for the cross sections from \citet{verner_etal:96}) and the uncertainty of the absolute calibration of TIMED/SEE. The uncertainty of the photoionization rates derived from the proxy is a sum of the uncertainty of the formula used to reconstruction and the systematic error hidden in the relation between the photoionization rates and the proxy. In this analysis we take into account only the uncertainty of the spread (standard deviation) of the daily data for each solar rotation average and the accuracy of the found relation between ionization rates and the proxy. The uncertainties do not include the uncertainty related to the accuracy of the spectral irradiance measured by TIMED. The error bars for the F10.7-based photoionization rates come from the uncertainty of the model, calculated from the formula for relative errors:
\begin{equation}
\delta_{\mathrm{model}} = \sqrt{\delta_{\mathrm{i,F10.7}}^2 + \delta_{\mathrm{fit}}^2},
\label{eq:DeltaPhModelSem}
\end{equation}
where $\delta_{\mathrm{i,F10.7}}$ is the standard deviation of the Carrington rotation averages of F10.7 used in the calculations expressed in percentage, and $\delta_{\mathrm{fit}}$ is the accuracy of the reconstruction of the photoionization rates from the derived model (Equation~\ref{eq:DeltaPhModelSem}), equal to $\sim5\%$. The spread of solar rotation averaged of F10.7 ranges from about $2\%$ in solar minimum to about $20\%$ in solar maximum, so the average accuracy of the proxy-based photoionization rates according Equation~\ref{eq:DeltaPhModelSem} is $\sim 12\%$ and ranges from less than $10\%$ during solar minimum up to $20\%$ during solar maximum. The relative uncertainty of the Carrington rotation averages of photoionization rates derived from direct integration of irradiance measured by TIMED is $\sim 2\% - 12\%$ for solar minimum and maximum, respectively.

The formula from Equation~\ref{eq:final} works only for the Carrington rotation averages of the F10.7 flux and is not recommended for use to reconstruct the daily rates of photoionization. The issues of reconstruction of the daily rates of ionization losses from proxies are described in details by \citet{bochsler_etal:14a}. In general, the features observed at short-timescales in both series (EUV flux in range relevant for helium ionization and F10.7 flux) come from distinct regions on the Sun and are weakly correlated with each other.

\section{Discussion and conclusions}
The reconstruction of the photoionization rates for the epoch when direct measurements of the solar EUV flux were not available can be done only from EUV proxies. The series of the EUV proxies needed for this should be as homogeneous in time as possible and free from calibration issues. Additionally, the reconstruction is only possible under the assumption that the relation between the investigated quantities (photoionization rates and the proxy, F10.7 in our case) does not change with time. Without a time series of the photoionization rates and the proxy longer than one solar cycle the validation of the possible secular changes of the solar irradiance is impossible. If such secular changes do exist, the reconstruction of the past photoionization rates from proxies is biased by the time range used to find the correlation.

Assuming that the correlation does not change with time, we updated the correlation formula between the photoionization rates of neutral interstellar He at 1~AU from the Sun calculated from the EUV irradiance directly measured by TIMED/SEE and the solar 10.7~cm flux, which is the longest available EUV proxy series, the most homogeneous (free from known intercalibration issues), stable, and continuously observed. 

We proposed a new method for reconstruction of the proxy-based ionization rates in the case when the distribution of data points between the minimum and maximum values is not uniform. We grouped the data into sectors based on the photoionization values and calculate averages of both quantities for each sector to use points of similar importance. We fit to the sectored data a nonlinear function $(y=ax^p+b)$ using an iterative algorithm with weights. The resulting value of the power index is $p \approx 0.57$. It appears almost a square root of the F10.7 flux, as noticed earlier by \citet{bzowski_etal:12b}. The reconstruction of the photoionization rates for He based only on F10.7 index gives rates about $10\%$ higher than the proxy-based model proposed by \citet{bochsler_etal:14a} and differ up to $15\%$ for solar maximum for years before TIMED launch and more than $20\%$ in the increasing phase of activity of SC~24 from the photoionization rates presented by \citet{bzowski_etal:13b}. The main reason for this difference seems to be rejecting the SEM measurements, which were shown by \citet{wieman_etal:14a} to be inaccurate. 

Recently \citet{didkovsky_wieman:14a} published an updated comparison of the EUV flux level from SEM measurements for solar minima. SEM is the only one time series of solar irradiance measured over two solar cycles. They confirm the decrease of the registered solar irradiance around HeII line in minimum of SC~23/24 in comparison to minimum of SC~22/23, which is about $12\%$.  Such a decrease is not present in the F10.7 series and observations in other solar radio wavebands, for which the base level for solar minimum is almost not cycle-dependent \citep[e.g.][]{dudokdewit_etal:14a, chen_etal:11a, svalgaard_hudson:10a}. This may suggest that the changes in the solar EUV are not fully reflected by the phenomena on the solar surface responsible for the F10.7 radio emission. This certainly would affect the photoionization rates derived from the F10.7 index, but the suggested difference between the minimum level for EUV is in the range of the uncertainty of the proposed proxy-based model.

\acknowledgments
This study is supported by the grant 2012/06/M/ST9/00455 from the Polish National Science Center. 
We acknowledge LASP for the release of TIMED/SEE data for public use and NOAA National Geophysical Data Center for F10.7 data release. 
	
\bibliographystyle{apj}

\end{document}